\documentclass{article}
\usepackage{spconf,amsmath,graphicx}
\DeclareMathOperator*{\argmax}{arg\,max}
\usepackage{booktabs,dcolumn,caption}
\usepackage{multirow}
\usepackage{amsfonts}
\usepackage{bbm}
\usepackage{url}
\usepackage[linesnumbered,ruled]{algorithm2e}

\title{A Large-Scale Deep Architecture for Personalized Grocery Basket Recommendations}
\name{\parbox{\linewidth}{\centering Aditya Mantha \qquad Yokila Arora \qquad Shubham Gupta \qquad Praveenkumar Kanumala \qquad Zhiwei Liu  Stephen Guo \qquad Kannan Achan }}
\address{Walmart Labs, Sunnyvale, California, USA}
\begin{document}
\ninept
\maketitle
\begin{abstract}
With growing consumer adoption of online grocery shopping through platforms such as Amazon Fresh, Instacart, and Walmart Grocery, there is a pressing business need to provide relevant recommendations throughout the customer journey. In this paper, we introduce a production within-basket grocery recommendation system, RTT2Vec, which generates real-time personalized product recommendations to supplement the user's current grocery basket. We conduct extensive offline evaluation of our system and demonstrate a 9.4\% uplift in prediction metrics over baseline state-of-the-art within-basket recommendation models. We also propose an approximate inference technique 11.6x times faster than exact inference approaches. In production, our system has resulted in an increase in average basket size, improved product discovery, and enabled faster user check-out.



\end{abstract}
\section{Introduction}
\label{sec:intro}

A critical component of a modern day e-commerce platform is a user-personalized system for serving recommendations. While there has been extensive academic research for recommendations in the general e-commerce setting, user personalization in the online groceries domain is still nascent. An important characteristic of online grocery shopping is that it is highly personal. Customers show both regularity in purchase types and purchase frequency, as well as exhibit specific preferences for product characteristics, such as brand affinity for milk or price sensitivity for wine. 


One important type of grocery recommender system is a within-basket recommender, which suggests grocery items that go well with the items in a customer's shopping basket, such as milk with cereals or pasta with pasta sauce. In practice, customers often purchase groceries with a particular intent, such as for preparing a recipe or stocking up for daily necessities. Therefore, a within-basket recommendation engine needs to consider both item-to-item compatibility within a shopping basket as well as user-to-item affinity, to generate efficient product recommendations that are truly user-personalized.

%
%




In this paper, we introduce Real-Time Triple2Vec,~\textbf{RTT2Vec}, a real-time inference architecture for serving within-basket recommendations. Specifically, we develop a representation learning model for personalized within-basket recommendation task, and then convert this model into an approximate nearest neighbour (ANN) retrieval task for real-time inference. Further, we also discuss some of the scalability trade-offs and engineering challenges when designing a large-scale, deep personalization system for a low-latency production application. 

 

For evaluation, we conducted exhaustive offline experiments on two grocery shopping datasets and observe that our system has superior performance when compared to the current state-of-the-art models. 
Our main contributions can be summarized as follows:
\begin{itemize}
    \item We introduce an approximate inference method which transforms the inference phase of a within-basket recommendation system into an Approximate Nearest Neighbour (ANN) embedding retrieval.
    \item We describe a production real-time recommendation system which serves millions of online customers, while maintaining high throughput, low latency, and low memory requirements.
\end{itemize}

\section{Related Work}
\label{sec:relatedwork}

Collaborative Filtering (CF) based techniques have been widely adopted in academia and industry for both user-item \cite{hu2008collaborative} and item-item recommendations \cite{linden2003amazon}. Recently,this approach has been extended to the within-basket recommendation task. The factorization-based models, \textbf{BFM} and \textbf{CBFM} \cite{le2017basket}, consider multiple associations between the user, the target item, and the current user-basket to generate within-basket recommendations. Even though these approaches directly optimize for task specific metrics, they fail to capture non-linear user-item and item-item interactions.

Due to the success of using latent representation of words (such as the \textbf{skip-gram} technique~\cite{mikolov2013distributed,mikolov2013efficient}) in various NLP applications, representation learning models have been developed across other domains. The \textbf{word2vec} inspired \textbf{CoFactor}~\cite{liang2016factorization} model utilizes both Matrix Factorization (MF) and item embeddings jointly to generate recommendations. \textbf{Item2vec}~\cite{barkan2016item2vec} was developed to generate item embeddings on itemsets. Using these, item-item associations can be modeled within the same itemset (basket). \textbf{Prod2vec} and \textbf{bagged-prod2vec}~\cite{grbovic2015commerce} utilize the user purchase history to generate product ads recommendations by learning distributed product representations. Another representation learning framework, \textbf{metapath2vec}~\cite{dong2017metapath2vec}, uses meta-path-based random walks to generate node embeddings for heterogenous networks, and can be adapted to learn latent representations on a user-item interaction graph. By leveraging both basket and browsing data jointly, \textbf{BB2vec}~\cite{trofimov2018inferring} learns dual vector representations for complementary recommendations. Even though the above skip-gram based approaches are used in wide areas of applications such as digital advertising and recommendation systems, they fail to jointly optimize for user-item and item-item compatibility.

\setlength{\belowcaptionskip}{-14pt}
\begin{figure*}[htb]
\begin{minipage}[b]{1.0\linewidth}
  \centering
  \centerline{\includegraphics[width=\textwidth,height=6cm]{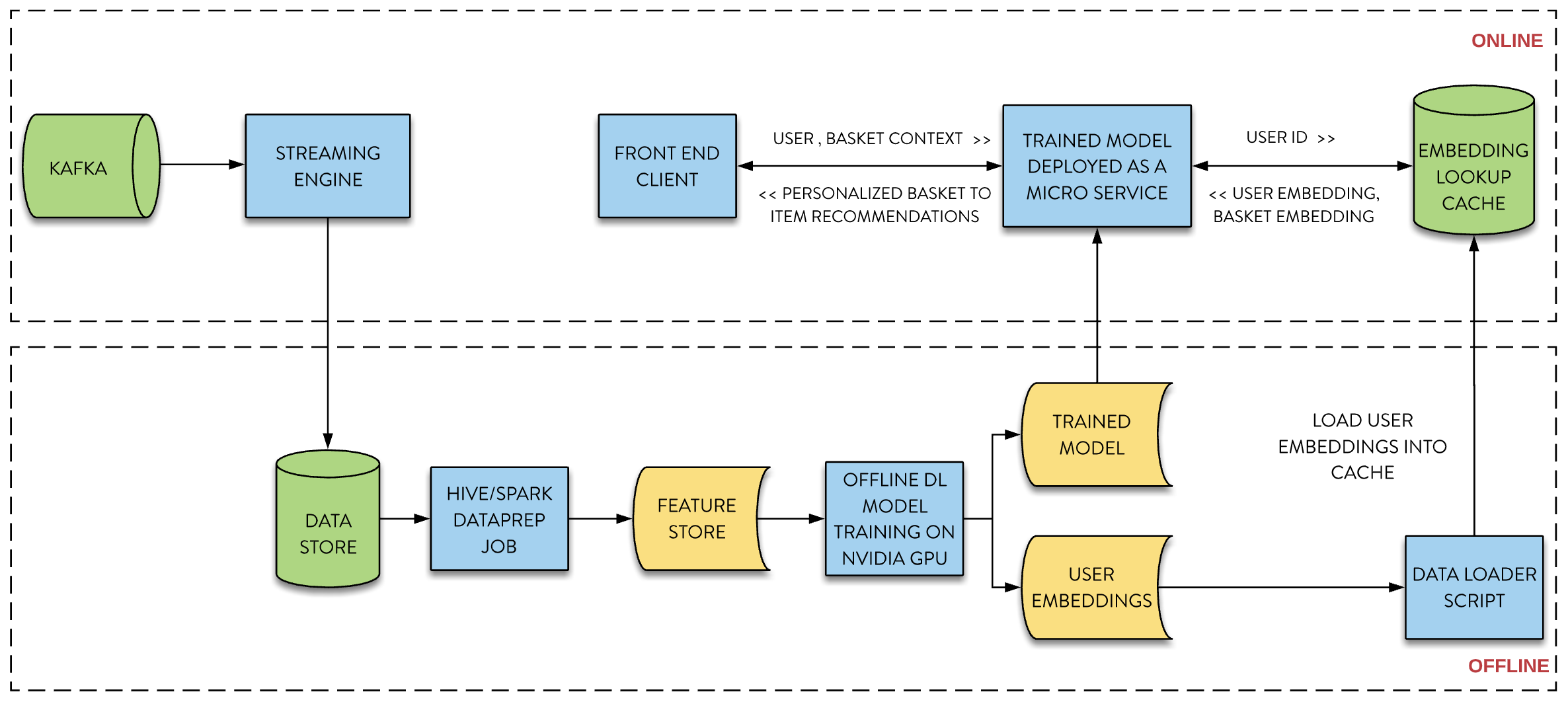}}
\end{minipage}
\caption{System Architecture for Real-Time Personalized Basket-to-Item Recommendations}
\label{fig:arch}
\end{figure*}

There has also been significant research to infer functionally complementary relations for item-item recommendation tasks. These models focus on learning compatibility \cite{veit2015learning}, complementarity \cite{zhang2018quality,kang2019complete,DBLP:journals/corr/abs-1904-12574}, and complementary-similarity \cite{mcauley2015inferring,mane2019complementary} relations across items and categories from co-occurrence of items in user interactions.


\section{Method}
\label{sec:method}


In this section, we explain the modeling and engineering aspects of a production within-basket recommendations system. First, we briefly introduce the state-of-the-art representation learning method for within-basket recommendation tasks, triple2vec. Then, we introduce our Real-Time Triple2Vec (RTT2Vec) system inference formulation, production algorithm, and system architecture.

\textbf{Problem Definition}: Consider $m$ users $\mathfrak{U}$ = $\{u_1, u_2, .....,u_m\}$ and $n$ items $\mathfrak{I}$ = $\{i_1,i_2,...i_n\}$ in the dataset. Let ${\mathfrak{B}}_u$ denote a basket corresponding to user $u \in \mathfrak{U} $, where basket refers to a set of items $\{i^{'} | i^{'} \in \mathfrak{I}\}$.  The goal of the within-basket recommendation task is given ($u$, ${\mathfrak{B}}_u$) generate top-k recommendations $\{i^{*} | i^{*} \in \mathfrak{I}\setminus {\mathfrak{B}}_u\}$ where $i^*$ is complementary to items in ${\mathfrak{B}}_u$ and compatible to user $u$.

\subsection{Triple2vec model}
\label{ssec:T2V}
We utilize the triple2vec~\cite{wan2018representing} model for generating personalized recommendations. The model employs (user $u$, item $i$, item $j$) triples, denoting two items ($i$, $j$) bought by the user $u$ in the same basket, and learns representation $h_u$ for the user $u$ and a dual set of embeddings ($p_i, q_j$) for the item pair ($i$, $j$).

\setlength{\belowdisplayskip}{0pt} \setlength{\belowdisplayshortskip}{0pt}
\setlength{\abovedisplayskip}{0pt} \setlength{\abovedisplayshortskip}{0pt}
\begin{equation} \label{cohesion_score}
\begin{split}
s_{i,j,u} = p_i^T q_j + p_i^T h_u + q_j^T h_u
\end{split}
\end{equation}

The cohesion score for a triple ($u,i,j$) is defined by Eq.~\ref{cohesion_score}. It captures both user-item compatibility ($p_i^T h_u $, $q_j^T h_u$) as well as item-item complementarity ($p_i^T q_j$). The embeddings are learned by maximizing the co-occurrence log-likelihood of each triple as:

\begin{equation} \label{likelihood}
\begin{split}
L = \sum_{\forall (i,j,u)}\log{P(i|j,u)}+\log{P(j|i,u)}+\log{P(u|i,j)}
\end{split}
\end{equation}
where $P(i|j,u)=\frac{\exp(s_{i,j,u})}{\sum_{i^{'}} \exp(s_{i^{'},j,u})}$. Similarly, $P(j|i,u)$ and $P(u|i,j)$ can be obtained by interchanging ($i$,$j$) and ($i$,$u$) respectively.


In accordance with most skip-gram models with negative sampling \cite{gutmann2012noise}, the softmax function in Eq.~\ref{likelihood} is approximated by the Noise Contrastive Estimation (NCE) loss function, using TensorFlow~\cite{abadi2016tensorflow}. A log-uniform (Zipf) distribution is used to sample negative examples.

\subsection{RTT2Vec: Real-Time Model Inference}
\label{ssec:RTI}
Serving a personalized basket-to-item recommendation system is challenging in practice. In conventional production item-item or user-item recommendation systems, model recommendations are precomputed offline via batch computation, and cached in a database for static lookup in real-time. This approach cannot be applied to basket-to-item recommendations, due to the exponential number of possible shopping baskets. Additionally, model inference time increases with basket size (number of items), making it challenging to perform real-time inference within production latency requirements.



\begin{equation} \label{argmax1}
\begin{split} 
\argmax_{j}(p_i^T q_j + p_i^T h_u + q_j^T h_u) =\argmax_{j}{(\underbrace{[p_i\quad h_u]}_\text{Query Vector}}^T \underbrace{[q_j\quad q_j]}_\text{ANN Index})
\end{split}
\end{equation}

We transform the inference phase of triple2vec (Section~\ref{ssec:T2V}) into a similarity search of dense embedding vectors. For a given user $u$ and anchor item $i$, this can be achieved by taking $argmax$ of the cohesion score (Eq.~\ref{cohesion_score}) and adjusting it as shown in Eq.~\ref{argmax1}. 
The first term, the \textbf{query vector}, depends on the inputs $u$ and $i$, and the second term, the \textbf{ANN index}, only depends on $j$, thus transforming our problem into a similarity search task.

\begin{equation} \label{argmax2}
\begin{split} 
\argmax_{j}(\frac{(p_i^T q_j + p_i^T h_u + q_j^T h_u) + (p_j^T q_i + p_j^T h_u + q_i^T h_u)}{2}) \\
= \argmax_{j}(p_i^T q_j  + q_j^T h_u + p_j^T q_i + p_j^T h_u) \\
= \argmax_{j}{(\underbrace{[p_i\quad h_u \quad q_i \quad h_u]}_\text{Query Vector}}^T \underbrace{[q_j \quad q_j \quad p_j \quad p_j]}_\text{ANN Index})
\end{split}
\end{equation}

Hashing is an effective and widely-used technique for fast large-scale data retrieval \cite{ye2019sequential,peng2019deep,zhang2017ssdh}. In accordance, we further speed up the similarity search of the inference problem by using an off-the-shelf Approximate Nearest Neighbour (ANN) indexing library, such as FAISS~\cite{JDH17}, ANNOY~\cite{annoy_lib}, or NMSLIB~\cite{naidan2016non,boytsov2013engineering}, to perform approximate dot product inference efficiently at large-scale. These ANN indexing schemes convert the traditional $O(nlog^{2}n)$ sort algorithm into a $O(log^{2}n)$ serial time\footnote{The complexity depends on the indexing scheme used. The reported complexity is achieved if we use FAISS as the indexing scheme.} algorithm by efficiently parallelizing it using multi-threading, BLAS and machine SIMD vectorization.

We also observe that model performance improves by interchanging the dual item embeddings and taking the average of the cohesion scores, as shown in Eq.~\ref{argmax2}.


\subsection{RTT2Vec: Production Algorithm}
\label{ssec:RRT2vec}


The RTT2Vec algorithm used for generating top-k within-basket recommendations in production consists of three principal tasks: basket-anchor set selection, model inference, and post-processing. These steps are described below in detail: 

\textbf{Basket-anchor set selection:} To generate personalized within-basket recommendations, we replace the item embeddings $p_i$ and $q_i$ with the average embedding of all the items in the shopping basket. This approach works very well for baskets with smaller sizes, but in practice, a typical family's shopping basket of groceries contains dozens of items. Taking the average of such large baskets results in losing information about the individual items in the basket. For larger baskets, we deploy a sampling algorithm which randomly selects 50\% of items in the basket as a basket-anchor set. 
 
\textbf{Model Inference:} For each item in the basket-anchor set, we create the query vector  $ [p_i\quad h_u \quad q_i \quad h_u ] $ using the pre-trained user embedding $h_u$ and item embeddings $p_i$ and $q_i$ (refer Eq.~\ref{argmax2}). Then, we search the query vector in the Approximate Nearest Neighbour (ANN) index to retrieve the top-k recommendations. 

The ANN index is created from the concatenation of the dual item embeddings $[q_j \quad q_j \quad p_j \quad p_j] \forall $ j $\in \mathfrak{I}$. The ANN index and embeddings are stored in memory for fast lookup. In practice, the inference can be further speed up by performing a batch lookup in the ANN index instead of performing a sequential lookup for each item in the basket-anchor set.

After the top-k recommendations are retrieved for each anchor item in the basket-anchor set, a recommendation aggregator module is used to blend all the recommendations together. The aggregator uses several factors such as number of distinct categories in the recommendation set, the individual item scores in the recommendations, taxonomy-based weighting, and business rules to merge the multiple recommendation sets, and filter to a top-k recommendation set. 

\textbf{Post-processing:} Once the top-k recommendation set is generated, an additional post-processing layer is applied. This layer incorporates diversification of items, removes blacklisted items and categories, utilizes market-basket analysis association rules for taxonomy-based filtering, and applies some business requirements to generate the final top-k recommendations for production serving.

\subsection{RTT2Vec: Production System Architecture}

In this section, we provide a high level overview of our production recommendation system as illustrated in Figure~\ref{fig:arch}. This system is comprised of both offline and online components. The online system consists of a memcached distributed cache, streaming system, a real time inference engine, and a front-end client. The offline system encompasses a data store, a feature store serving all the recommendation engines at Walmart, and an offline model training framework deployed on a cluster of GPUs.

At Walmart Grocery, we deal with a large volume of customer interactions, streaming in at various velocities. We use the Kafka streaming engine to capture real-time customer data without delay and store the data in a Hadoop-based distributed file system. For offline model training, we construct training examples by extracting features from our feature store through Hive and Spark jobs. Then, the training examples are input into an offline deep learning model, which is trained on a GPU cluster, generating user and dual-item embeddings. These embeddings are then stored in an embedding store (distributed cache) to facilitate online retrieval by the real-time inference engine. 

The primary goal of deploying a real-time inference engine is to provide personalized recommendations, while ensuring very high throughput and providing a low-latency experience to the customer. The real-time inference engine utilizes a Approximate Nearest Neighbor (ANN) index, constructed from the trained embeddings, and deployed as a micro-service. This engine interacts with the front-end client to obtain user and basket context and generates personalized within-basket recommendations in real-time.

\section{Experiments}
\label{sec:exp}

\subsection{Datasets}

Our experimental evaluation is performed on one public dataset and one proprietary dataset. Both datasets are split into train, validation, and test sets. The public Instacart dataset is already split into prior, train and test sets. For the Walmart Grocery dataset, the train, validation, and test sets comprise of one year, the next 15 days, and the next one month of transactions respectively. 

\begin{itemize}
\item \textbf{Instacart:} We use the open-source grocery dataset published by Instacart~\cite{instacart_data}, containing approximately 206k users and 50k items
with 3.4m total interactions. The average basket size is 10. 

\item \textbf{Walmart:} We use a subset of a proprietary online Walmart Grocery~\cite{walmart_grocery} dataset for these experiments. The dataset contains approximately 3.5m users and 90k items with 800m interactions.
\end{itemize}
\addtolength{\parskip}{-0.5mm}
\begin{table} 
\centering
\vspace{-1em}
\caption{Within-Basket Recommendations}
\vspace{-1em}
\label{tab:within-basket}
\begin{tabular}{clcc}
\toprule
 \textbf{Dataset}& \textbf{Method} & \textbf{Recall@20} & \textbf{NDCG@20}  \\ 
\midrule
\multirow{6}{*}{\textbf{Instacart}} & ItemPop      & 0.1137  & 0.1906 \\
    & BB2vec              & 0.0845    & 0.1258 \\
    & item2vec      & 0.0810 & 0.1356  \\
    & triple2vec (NP)             & 0.0794 & 0.1709 \\
    & triple2vec & 0.1354*   & 0.1876*  \\
    & RTT2Vec & \textbf{0.1481}  &   \textbf{0.2391} \\
    \midrule
    & Improv.\% & $\textbf{9.37\% }$ & $\textbf{21.53\%}$\\
    \midrule
  \multirow{6}{*}{\textbf{Walmart}} & ItemPop      & 0.0674  & 0.1318  \\
    & BB2vec        & 0.0443   & 0.0740  \\
    & item2vec          & 0.0474 & 0.0785  \\
    & triple2vec (NP)   & 0.0544 &  0.0988 \\
    & triple2vec        &  0.0685*  &  0.1142*  \\
    & RTT2Vec           & \textbf{0.0724}  & \textbf{0.1245} \\
    \midrule
    & Improv.\% & $\textbf{5.75\% }$ & $\textbf{9.01\%}$\\
\bottomrule
\end{tabular}
\label{table:1}
\end{table}
\subsection{Evaluation}


\textbf{Metrics}: We evaluate the performance of models with the metrics: \textbf{Recall@K} and \textbf{NDCG@K}. \textbf{Recall@K} measures the fraction of relevant items successfully retrieved when the top-K items are recommended. \textbf{NDCG@K} (Normalized Discounted Cumulative Gain) is a ranking metric which uses position in the recommendation list to measure gain. Metrics are reported at K=20.

For the within-basket recommendation task, given a subset of the basket, the goal is to predict the remaining items in the basket. Let the basket be split into two sets $B_T$ and $B_{T'}$, where $B_T=\{i_1, i_2, .....,i_m\}$ denotes the subset of items in basket used for inference, and $B_{T'}=B \setminus{B_T} = \{j_1, j_2, .....,j_n\}$ denotes the remaining set of items in the basket. Let $S_K=\{r_1, r_2, .....,r_K\}$ denote the top-K recommendation list generated using $B_T$. Then:

\begin{equation}\label{recall_k}
    {\rm Recall@K} = \frac{|S_K \cap B_{T'}|}{|B_{T'}|}
\end{equation}

\begin{equation}\label{ndcg_k}
    {\rm NDCG@K} =  \sum_{p=1}^{k} \frac{\mathbbm{1}{[l_p \in B_{T'}]}}{\log_{2}{(p+1)}} 
\end{equation}

where $p$ denotes the rank of the item in the recommended list $l$, and $ \mathbbm{1}{}$ is the indicator function indicating if $l_p \in B_{T'}$. 

\setlength{\belowcaptionskip}{-17pt}
\begin{figure}[htb]
\begin{minipage}[b]{1.0\linewidth}
  \centering
  \centerline{\includegraphics[width=\textwidth]{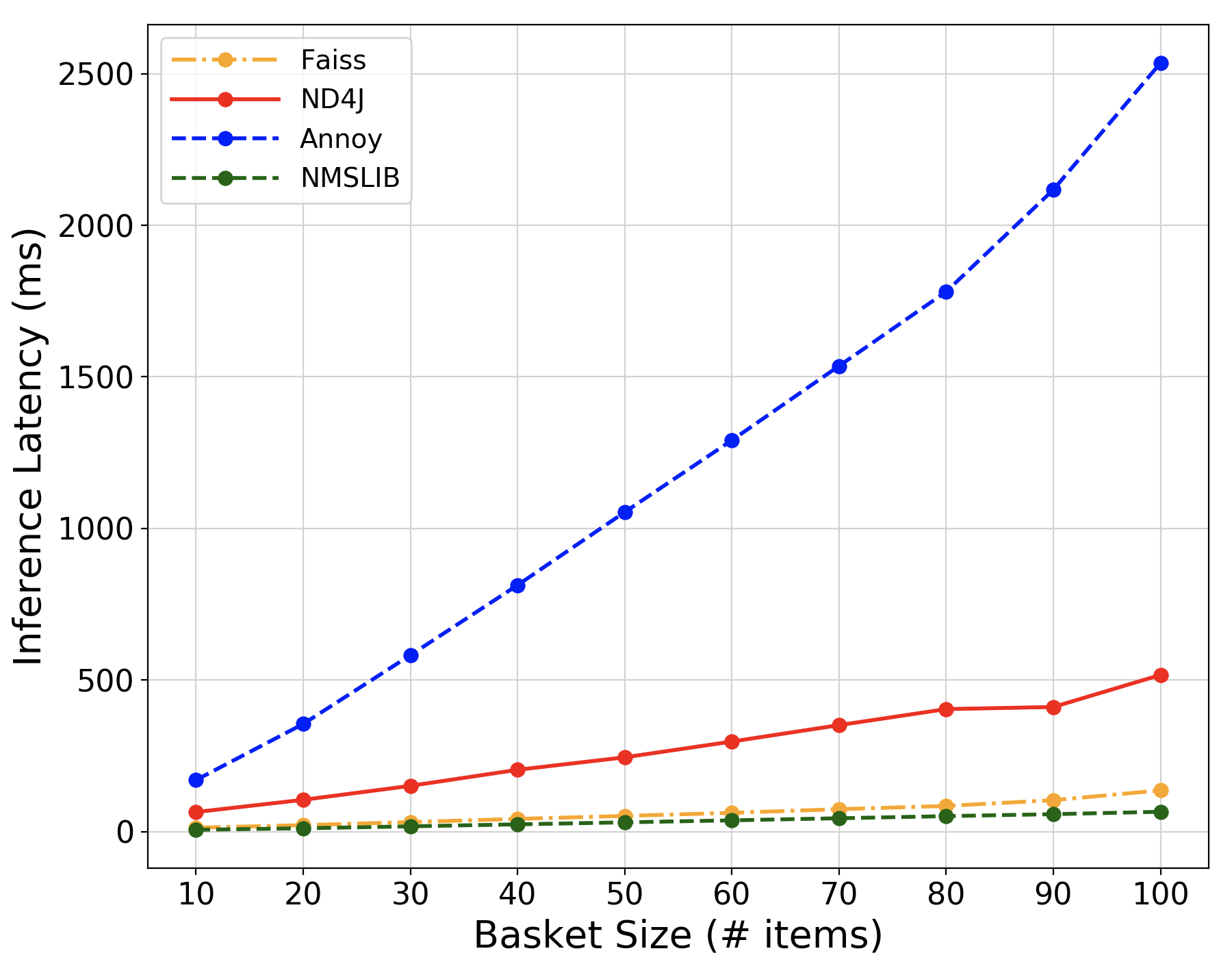}}
\end{minipage}
\caption{System Latency Comparison}
\label{fig:latencygraph}
\end{figure}

\subsection{Baseline Models}
\label{baseline_models}
Our system is evaluated against the following models:
\begin{itemize}
  \item \textbf{ItemPop}: The Item Popularity (ItemPop) model selects the top-K items based on their frequency of occurrence in the training set. The same set of items are recommended for each test basket for each user. 
    \item \textbf{item2vec}: The item2vec \cite{barkan2016item2vec} model uses Skip-Gram with Negative Sampling (SGNS) to generate item embeddings on itemsets. We apply this model on within-basket item-sets to learn co-occurrence of items in the same basket. 
 \item \textbf{BB2vec}: The BB2vec \cite{trofimov2018inferring} model learns vector representations from basket and browsing sessions. For a fair comparison with other models, we have adapted this method to only use basket data and ignore view data.
  \item \textbf{triple2vec (NP)}: This is a non-personalized variation of triple2vec (as explained in Section \ref{ssec:T2V}), where we only use the dual item embeddings and ignore user embeddings during inference. The cohesion score here (Eq. \ref{cohesion_score}) can be re-written as: $s_{i,j} = p_i^T q_j$
  \item \textbf{triple2vec}: The state-of-the-art triple2vec model (as explained in Section \ref{ssec:T2V}) employs the Skip-Gram model with Negative Sampling (SGNS), applied over (user, item, item) triples in the test basket to generate within-basket recommendations.
\end{itemize}

\textbf{Parameter Settings}:  We use an embedding size of 64 for all skip-gram based techniques, along with the Adam Optimizer with a initial learning rate of 1.0, and the noise-contrastive estimation (NCE) of softmax as the loss function. A batch size of 1000 and a maximum of 100 epochs are used to train all skip-gram based models. We use 5 million triples to train the Instacart dataset and 200 million triples for the Walmart dataset.

\subsection{Results}

We next evaluate our model predictive performance and system latency.
The models are trained on an NVIDIA K80 GPU cluster, each consisting of 48 CPU cores. For evaluation and benchmarking, we use an 8-core x86\_64 CPU with 2-GHz processors.

\textbf{Predictive Performance}:
We compare the performance of our system, RTT2Vec, against the models described in Section~\ref{baseline_models} on the within-basket recommendation task using NMSLIB. For each basket in the test set, we use 80\% of the items as input and  the remaining 20\% of items as the relevant items to be predicted. As displayed in Table~\ref{table:1}, we observe that our system outperforms all other models on both Instacart and Walmart datasets, improving Recall@20 and NDCG@20 by 9.37\% (5.75\%) and 21.5\% (9.01\%) for Instacart (Walmart) datasets when compared to the current state-of-the-art model triple2vec. 



\textbf{Real-Time Latency}: Further, we test real-time latency for our system using exact and approximate inference methods as discussed in Section~\ref{sec:method}. Figure~\ref{fig:latencygraph} displays system latency (ms) versus basket size. To perform exact inference based on Eq.~\ref{argmax2}, we use ND4J~\cite{nd4j} and for approximate inference (as discussed in Section~\ref{ssec:RTI}), we test Faiss, Annoy, and NMSLIB libraries. 

ND4J is a highly-optimized scientific computing library for the JVM. Faiss is used for efficient similarity search of dense vectors that can scale to billions of embeddings, Annoy is an approximate nearest neighbour library optimized for memory usage and loading/saving to disk ,and NMSLIB is a similarity search library for generic non-metric spaces. 

On average, ND4J adds 186.5ms of latency when performing exact real-time inference. For approximate inference, Faiss, Annoy, and NMSLIB libraries add an additional 29.3ms, 538.7ms, and 16.07ms of system latency respectively. Faiss and NMSLIB provide an option to perform batch queries on the index, therefore latency is much lower than Annoy. Faiss and NMSLIB are 6-10 times faster than the exact inference method using ND4J. In practice, we use NMSLIB in our production system as it provides better overall performance. 
 
\section{Conclusion and Future Work}
\label{sec:conclusion}
In this paper, we propose a state-of-the-art real-time user-personalized within-basket recommendation system, RTT2vec, to serve personalized item recommendations at large-scale within production latency requirements. As per our knowledge, this study is the first description of a large-scale production grocery recommendation system in the industry.
Our approach outperforms all baseline models on evaluation metrics, while respecting low-latency requirements when serving recommendations at scale.

Due to the increasing adoption of online grocery shopping and the associated surge in data size, there is an increase in the training time required for deep embedding models for personalized recommendations. Future work includes investigating the performance tradeoff of different sampling methodologies during model training. We are also exploring the introduction of additional content and contextual embeddings for improving model predictions further.  



\bibliographystyle{IEEEbib}
\bibliography{strings,refs}

\end{document}